%% file: main.tex
\documentclass[10pt, conference, compsocconf]{IEEEtran}

\IEEEoverridecommandlockouts
\usepackage{cite}
\usepackage{amsmath,amssymb,amsfonts}
\usepackage{algorithm}
\usepackage{algpseudocode}
\usepackage{graphicx}
\usepackage{textcomp}
\usepackage{xcolor}
\usepackage{caption}
\usepackage{subcaption}
\usepackage{multirow}
\usepackage{balance}
\usepackage{nopageno}
\usepackage{amssymb} % For checkmark symbol
\usepackage{adjustbox} % For \rot command to rotate text
\usepackage{array}
\usepackage[left=1.62cm,right=1.62cm,top=0.72in]{geometry}

\def\BibTeX{{\rm B\kern-.05em{\sc i\kern-.025em b}\kern-.08em
    T\kern-.1667em\lower.7ex\hbox{E}\kern-.125emX}}
\begin{document}

\title{IPBAC: Interaction Provenance-Based Access Control for Secure and Privacy-Aware Systems
~\vspace{-12pt}
}

\author{

\IEEEauthorblockN{\textsuperscript{} Sharif Noor Zisad and Ragib Hasan}
\IEEEauthorblockA{\textit{Department of Computer Science, University of Alabama at Birmingham} \\
Birmingham, AL 35295, USA \\
\{szisad,ragib\}@uab.edu\\
}
~\vspace{-32pt}
}

\maketitle

\thispagestyle{plain}
\pagestyle{plain}

\input{introduction}
\input{interaction_provenance}
\input{result}
\vspace{-10pt}
\balance
\bibliographystyle{IEEEtran}
\bibliography{references}

\end{document}

%% file: introduction.tex
\section{Introduction}
\label{sec:introduction}
\vspace{-5pt}
Interaction provenance refers to the documentation of every action and interaction within a system, including detailed metadata such as the actor's identity, the time the action occurred, and the surrounding context of the interaction~\cite{khan2015fuzzy}. This concept is important for understanding the history and origins of data and processes, providing a transparent and traceable record of all activities within a system specially privacy-aware systems.

Access control is a security measure that restricts access to resources within a system, ensuring that only authorized users can perform specific actions or access certain data \cite{sandhu1994access}. Traditional access control methods, such as Role-Based Access Control (RBAC)~\cite{sandhu1998role}, Attribute-Based Access Control (ABAC)~\cite{hu2015attribute}, and History-based access control (HBAC)~\cite{edjlali1998history}, each offer distinct approaches to managing access.

The integration of fuzzy logic~\cite{hajek2013metamathematics} with interaction provenance represents a significant advancement in addressing the limitations of traditional access control mechanisms. While conventional methods often lack the flexibility to adapt to complex, evolving contexts, fuzzy logic introduces a way to make decisions that consider uncertainties and gradations. By combining interaction provenance, it becomes possible to implement dynamic and context-aware policies that align closely with real-world scenarios. This integration allows decisions to move beyond binary outcomes, granting access with confidence levels based on historical interactions and user behavior.

% In this framework, fuzzy logic assigns scores to factors such as user reliability, contextual relevance, and historical engagement, which are processed using membership functions~\cite{deng2021information} and fuzzy rules~\cite{angelov2012new}. These scores are then used to determine whether access should be granted, partially allowed, or denied. Such a system ensures that only authorized actions are performed, while simultaneously enabling rigorous auditing and compliance through the interaction logs. This approach bridges the gap between traditional access control processes, whether human-to-human, human-to-machine, or machine-to-machine. Integrating fuzzy logic with IPBAC not only enhances security but also ensures traceability and fairness in highly dynamic and sensitive environments.

In this framework, fuzzy logic assigns scores to factors such as user reliability, contextual relevance, and historical engagement, which are processed using membership functions~\cite{deng2021information} and fuzzy rules~\cite{angelov2012new}. These scores are then used to determine whether access should be granted, partially allowed, or denied. Such a system ensures that only authorized actions are performed, while simultaneously enabling rigorous auditing and compliance through the interaction logs. This approach bridges the gap between traditional access control processes, whether human-to-human, human-to-machine, or machine-to-machine, by providing a comprehensive and adaptable decision-making process. Integrating fuzzy logic with IPBAC not only enhances security but also ensures traceability and fairness in highly dynamic and sensitive environments.

% By combining interaction provenance, which provides a detailed record of all actions and interactions, with fuzzy logic, 

%% file: interaction_provenance.tex
% \vspace{-20pt}
\section{Interaction Provenance-based Access Control}
\label{sec:interaction_provenance}
% \vspace{-5pt}
\subsection{Working Principal}
The terminologies of interaction provenance are as follows.

\subsubsection{Principal} A Principal is any party or entity involved in an interaction provenance record, having either transmitted or received a message or initiated an action.

\subsubsection{Event} An Event is a documented occurrence of a protocol or service execution that happened in the past, presumed to be initiated by a principal during an interaction.

\subsubsection{Interaction} The Interaction for a given event or series of events is represented by an ordered sequence of messages or actions exchanged between two or more entities.

\subsubsection{Interaction Provenance} The Interaction Provenance of a principal is a tamper-resistant, chronologically ordered record of interactions and events. In service-oriented computing, interactions are fundamentally driven by specific events or sets of events. At any point in time, a principal can only interact with another principal based on a single event, given that time follows a linear progression. This leads to the creation of a chronologically ordered chain of interaction provenance for each principal. Interactions are uniquely tied to a principal and cannot be transferred to others.

\input{system_architecture}

%% file: system_architecture.tex
\vspace{-4pt}
\subsection{System Architecture}
\label{sec:system_architecture}
\begin{figure}[!ht]
    \includegraphics[width=1\columnwidth]{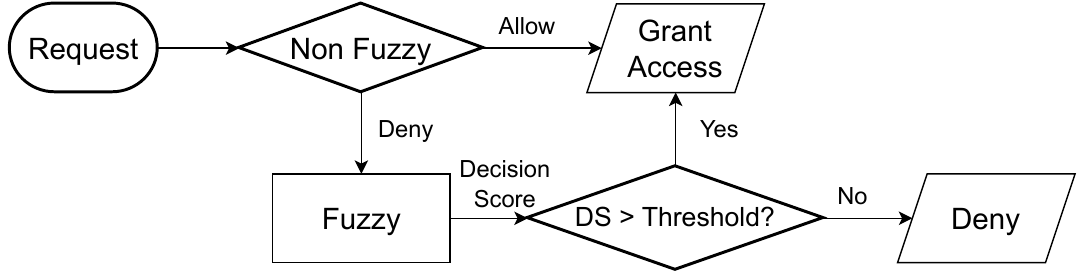}
    \vspace{-10pt}
    \centering
    \caption{System Flow Chart}
    \label{fig_system_flow_chart}
    \vspace{-10pt}
\end{figure}

\vspace{-2pt}
In the Interaction Provenance-Based Access Control (IPBAC) model, several key components perform significant roles, ensuring efficient operation and security. The flow chart of the system is depicted in Figure \ref{fig_system_flow_chart}.

\vspace{-10pt}

The request is first sent to the Non-Fuzzy model for access. The model then evaluates the request based on the previous log. If it allows for the request, the system grants access. Otherwise, the request is being sent to the Fuzzy model. The fuzzy model determines access by calculating a Decision Score (DS) derived from the user's previous successful interactions stored in the system. If the DS exceeds a predefined threshold, partial access is granted; otherwise, it is denied. 
Partial access means granting a limited subset of permissions. For example, users with moderate reliability might receive read-only access to sensitive incident data without permissions to modify or delete records.
% Partial access means a user can do some tasks but not others, based on rules or experience. For instance, an employee with a strong track record may be permitted to view sensitive reports but not edit or delete them.
This approach mirrors real-world scenarios, where trust and access privileges are granted based on an individual’s contributions and reliability within an organization. By incorporating interaction history, the model ensures that decisions are context-aware and reflective of a user's role. The access validation process using IPBAC is presented in Table \ref{tab_access_control}. 

\vspace{-5pt}
\begin{table}[!ht]
   \caption{Access Validation Process}
    \vspace{-2pt}
    \centering
    \begin{tabular}{|p{2.5cm}|p{5.2cm}|}
    \hline    
    \textbf{Step} & \textbf{Description} \\
    \hline
    User Request & The user initiates an action, sending a request to perform a task within the system. \\
    \hline
    Check Interaction Logs & The system checks historical interaction logs to see the user's previous actions, roles, and context. \\
    \hline
    Check Policies & The system simultaneously retrieves predefined access control policies related to the requested task. \\
    \hline
    Evaluate Role & The system evaluates the user's current role, previous interactions (from logs), and policy rules to assess if the request can proceed. \\
    \hline
    Deny Access & If the logs or policies do not allow the action, the system denies access, preventing further action. \\
    \hline
    Allow Access & If the user's role and logs meet the necessary policy conditions, the system grants access, allowing the action. \\
    \hline
    \end{tabular}
    \label{tab_access_control}
    \vspace{-8pt}
\end{table}

%% file: result.tex
\section{EXPERIMENTAL ANALYSIS}
\label{sec:result}

\begin{figure}[!ht]
    \includegraphics[width=0.8\columnwidth]{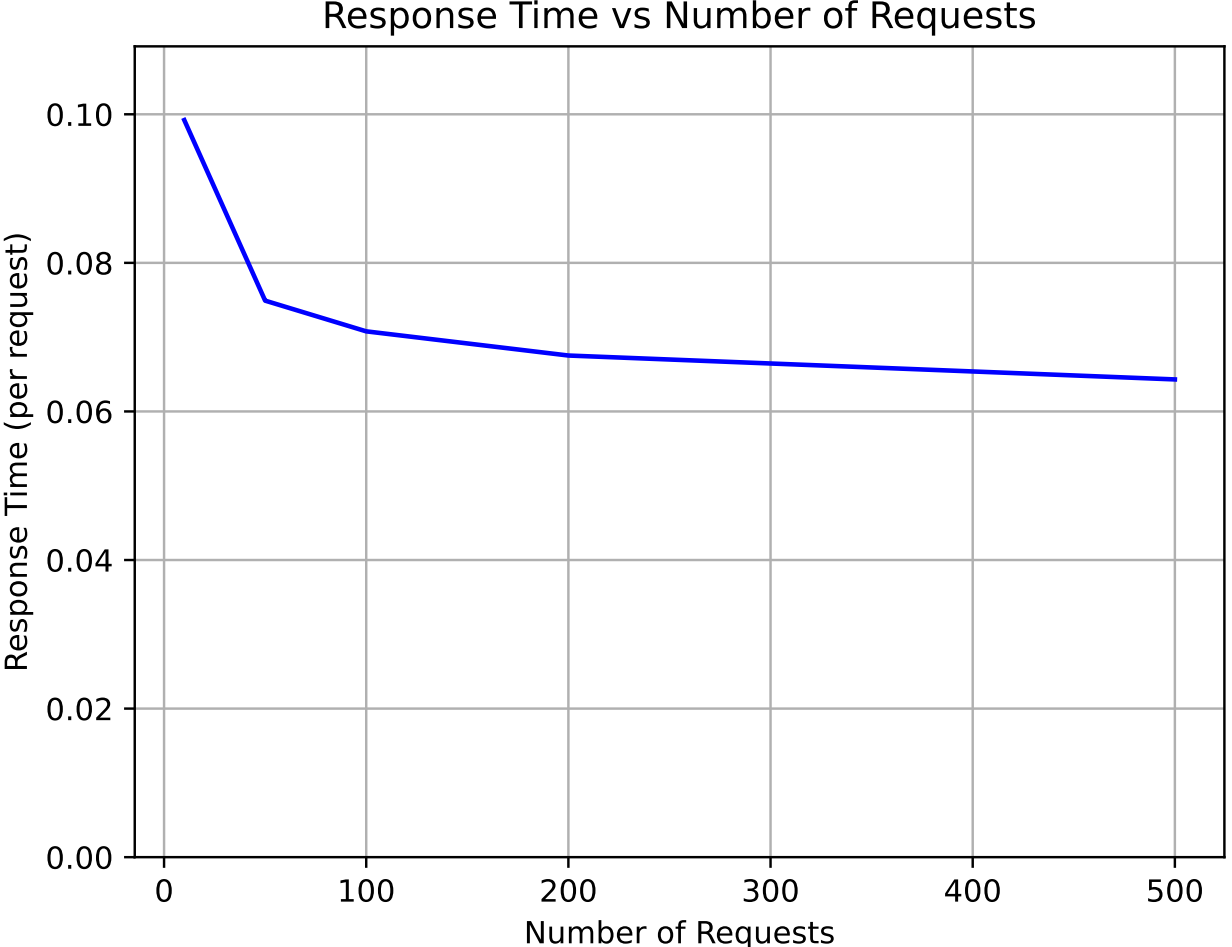}
    \centering
  %  \vspace{-15pt}
    \caption{Performance of the system}
    \label{fig_response_time}
   \vspace{-10pt}
\end{figure}

Performance of the system is measured using response time, which is described in Figure \ref{fig_response_time}.

\vspace{-8pt}

The response time plot in Figure \ref{fig_response_time} shows a relatively stable response time across all requests, with minimal fluctuations, indicating that the system can process requests efficiently without significant delays, even as the load increases. There is a slight decrease initially, but after around 100 requests, the response time stabilizes at approximately 0.065 seconds per request. This suggests that the system handles increasing requests efficiently without significant delays.

\begin{figure}[!ht]
    \includegraphics[width=0.8\columnwidth]{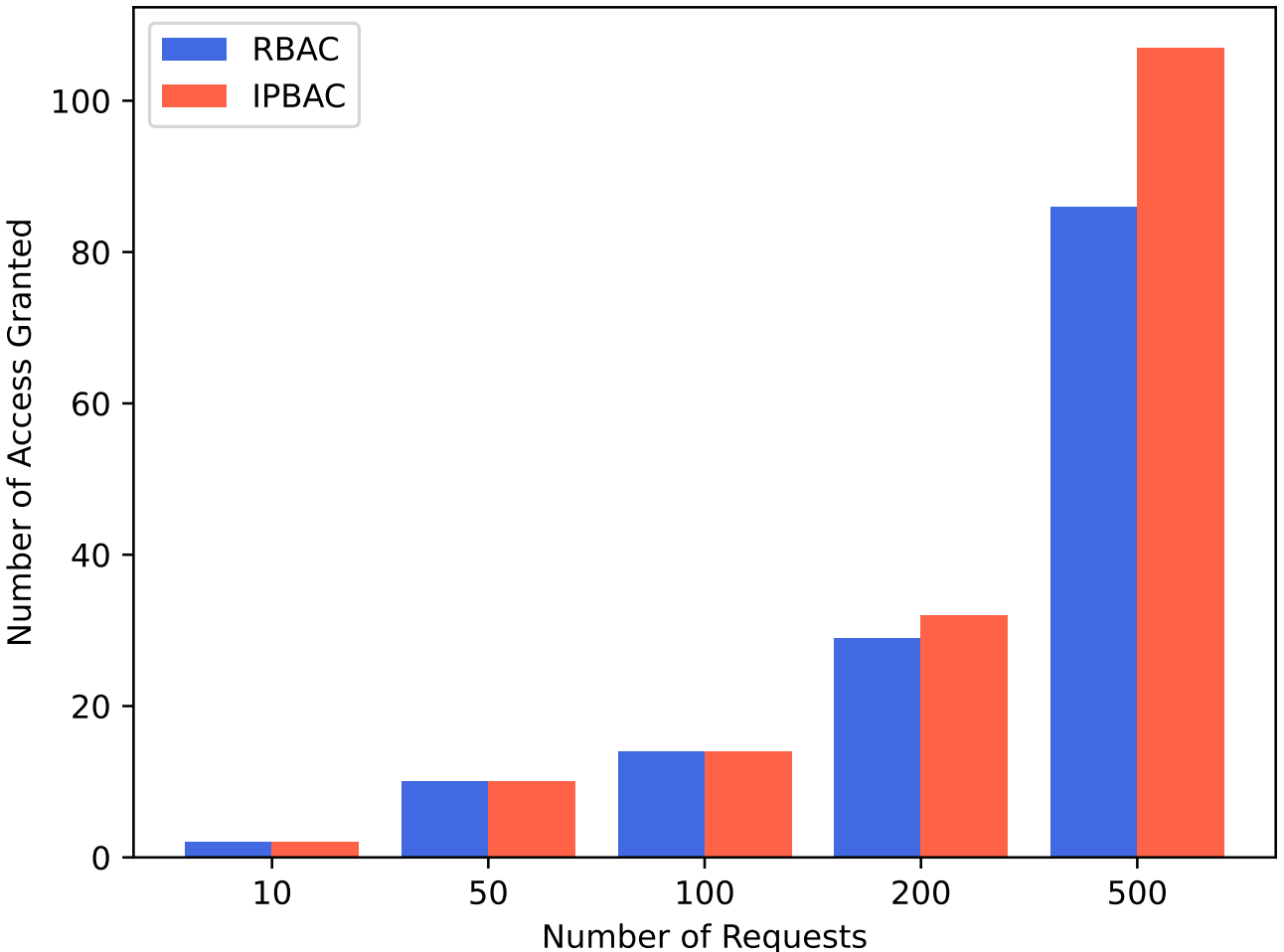}
    \centering
   % \vspace{-15pt}
    \caption{RBAC vs IPBAC Access Comparison}
    \label{fig_access_comparison}
    \vspace{-8pt}
\end{figure}

\vspace{-2pt}
We compared our IPBAC system with the existing RBAC (Role-Based Access Control) system. During testing of the system, there were almost 15,000 interaction provenances. In this comparison, the DS threshold ($\alpha$) was 0.2645 as there was a smaller number of interaction logs. It can be dynamically adjusted via periodic review of decision outcomes and their correctness, ensuring optimal balance between permissiveness and security. We provide full access ($\theta = 1$) in the fuzzy decision function instead of partial access to have a better comparison with RBAC. Figure \ref{fig_access_comparison} compares the number of access requests granted under RBAC and IPBAC as the number of requests increases from 10 to 500.

\vspace{-10pt}

The bar graph in Figure \ref{fig_access_comparison} shows that as the number of requests increases from 10 to 500, the number of granted access requests also increases for both RBAC and IPBAC. At 10 requests, both models grant 2 accesses. As requests increase to 50, both grant 10 accesses, showing no difference at this stage. However, at 100 requests, both grant 14 accesses, but as the number of requests grows, IPBAC starts granting more access than RBAC. At 200 requests, IPBAC grants 32 accesses, while RBAC grants 29, showing a slight advantage. This difference becomes more significant at 500 requests, where IPBAC grants 107 accesses, compared to RBAC’s 86. This suggests that IPBAC adapts better to increasing requests, making it more flexible in granting access. While RBAC enforces static role-based permissions, IPBAC leverages interaction provenance, allowing for more context-aware decision-making. This can be particularly valuable in systems where access needs to be fine-tuned based on past interactions rather than rigid roles.\\